\documentclass[prb,showpacs,twocolumn,superscriptaddress,10pt]{revtex4-2}

\usepackage{amssymb}
\usepackage{graphicx}
\usepackage{bm}
\usepackage{xcolor}
\usepackage{epsfig}
\usepackage{amsfonts}

\newcommand{\beqa}{\begin{eqnarray}}
\newcommand{\beeq}{\begin{equation}}
\newcommand{\eeqa}{\end{eqnarray}}
\newcommand{\eeqe}{\end{equation}}

\begin{document}

\title{Quantum State Transfer in a Magnetic Atoms Chain Using a Scanning Tunneling Microscope}

\author{Acosta Coden Diego Sebastian}

\affiliation{Instituto de Modelado e
Innovaci\'on Tecnol\'ogica (CONICET-UNNE), Avenida Libertad 5400, W3404AAS
Corrientes, Argentina.}

\affiliation{Facultad de Ciencias Exactas, Naturales y Agrimensura, Universidad Nacional del Nordeste, Avenida Libertad 5400, W3404AAS
Corrientes, Argentina.}

\author{Osenda Omar}

\affiliation{Instituto de Física Enrique Gaviola y Facultad de Matemática 
Astronomía y Física, Córdoba, Argentina}

\author{Ferr\'on Alejandro} 
\affiliation{Instituto de Modelado e
Innovaci\'on Tecnol\'ogica (CONICET-UNNE), Avenida Libertad 5400, W3404AAS
Corrientes, Argentina.}

\affiliation{Facultad de Ciencias Exactas, Naturales y Agrimensura, Universidad
Nacional del Nordeste, Avenida Libertad 5400, W3404AAS
Corrientes, Argentina.}

\begin{abstract}
The electric control of quantum spin chains has been an outstanding goal for the few last years due to its potential use in technologies related to quantum information processing. 
In this work, we show the feasibility of the different steps necessary to perform  controlled quantum state transfer in a $S=1/2$ titanium atoms chain employing the electric field produced by a Scanning Tunneling Microscope (STM). Our results show that the initialization and transmission of a single excitation state is achievable in short times, and with high fidelity. Our study uses spin Hamiltonians to model the magnetic atoms chain, the tip of the STM, the interaction between it and the atoms chain and the electronic response to the fields applied by the tip, employing sets of parameters compatible with the latest experiments and ab initio calculations.  The time dynamical evolution is considered in the full Hilbert space and the control pulses frequencies exerted by the tip of the microscope are within the reach of present day technology. 
\end{abstract}
\date{\today}

\maketitle

\section{Introduction}

The design of channels for the efficient transmission of quantum states is a topic of enormous importance in the present, with important implications for the future of quantum technologies. A compelling idea is to employ atomic-level fabrication using a scanning tunneling microscope (STM) \cite{Baumann2015, yang2017, yang2021resvalence,phark2023double, wang2023qb}, which allows precise atomic manipulation on a surface. This cutting-edge technology enables meticulous atomic manipulation on a surface with great precision. The distinctive advantage of employing an STM is not only related to its capacity to construct the desired structures but also to the ability to utilize a radio-frequency (RF) voltage for the coherent control of the electron spins associated with adatoms, that is, magnetic atoms adsorbed in non-magnetic surfaces \cite{chen2023spinchain}. Furthermore, we can read the spin states of these atoms through a spin-polarized tunnel current \cite{Baumann2015}.

Quantum state transfer, or transmission in arrays of qubits, is a subject with a long history \cite{Bose2003}. The transfer channel consists of a physical system comprising several copies of the same unit in the case of homogeneous systems or different units in the case of heterogeneous ones \cite{Matsukevich2004, Guccione2020, Davanco2017}. The transmission takes a quantum state prepared on a set of units to another set of units. In the simplest case, from a single initial unit to a single final unit. In actual implementations, the units could be superconductor qubits such as transmons \cite{Koch2007, Barends2014}, quantum dots with one or several electrons trapped inside \cite{Kandel2021}, cold atoms \cite{Yang2016, Murmann2015, Lorenzo2017}, nuclear spins \cite{Vandersypen2001}, effective two-level systems such as NV centres \cite{Gulka2021}, or adatoms. The effective Hamiltonians of the listed systems are spin Hamiltonians, and the interactions between the components of the systems determine the interaction terms in the spin Hamiltonian. In some cases, the effective Hamiltonian resembles a set of coupled oscillators, where the coupling term stands for the interaction between the actual units. 

Notwithstanding the physical features of the many microscopic and nanoscopic systems already considered in the literature as viable qubits, once the effective Hamiltonian is available, the quality of the corresponding transmission channel depends on its geometry layup, the number of elements (in the case of a chain its length),  the transmission protocol, the presence or not of actuators controlling the dynamics and many other factors. Although numerous systems show theoretical and practical high-quality quantum state transfer, the search for novel systems and control mechanisms continues. 

In the last decades, the STM tip has enabled the construction of various atomic arrangements. Based on the pioneering work of Baumann \cite{Baumann2015}, where  Electron Spin Resonance with STM (ESR-STM) \cite{yang2017,phark2023double,soohion2023b,sooyhon2023,wang2023qb,reale2024Er,chen2023spinchain} was developed,
there was an increase in interest in designing atomic-arranged configurations to perform diverse quantum control operations. Over the past nine years, researchers have successfully deposited Fe atoms on MgO at the Oxygen position and positioned Ti atoms at various locations on MgO. Isolated atoms deposited on different surfaces allow, using ESR-STM, precise control of electronic spins and measurement of properties such as the hyperfine coupling \cite{willke2018}. Additionally, in a recent experiment, the nuclear spin of a Cu atom on MgO was controlled \cite{yang2018}. As the experiments with isolated atoms progressed, attempts began to reproduce them in slightly more complex structures. The first ESR-STM experiment on Ti dimers was carried out in 2017 and reported in the work of Yang and collaborators \cite{yang2017}. After this, various experiments have been carried out with small arrangements, mainly made of Ti atoms, among which we can highlight one with arranges of four Ti atoms \cite{yang2021resvalence} and another studying the coherent evolution of two coupled Ti atoms \cite{veldman21}. Of particular interest are the latest experiments with heterogeneous arrangements \cite{phark2023double,soohion2023b,sooyhon2023,wang2023qb,reale2024Er}. Remote control of Ti atoms has been achieved by assembling Ti dimers to a Fe atom \cite{phark2023double}, leading to the design of the first coupled qubits using magnetic atoms \cite{wang2023qb}. In Wang {\it et. al.} \cite{wang2023qb} the authors demonstrate atom-by-atom construction using Ti and Fe atoms, coherent control, and readout of coupled electron-spin qubits using STM. The challenge of creating larger arrays and implementing control tasks on these devices will dominate part of the field in the coming years.

In this work, we propose a novel methodology for transmitting states in a spin chain made up of magnetic atoms using an STM.  Here, we model a chain of $S=1/2$ \cite{yang2017,willke2018,ferron2019} Ti atoms coupled to an Fe atom, which acts as the STM tip\cite{lado2017,Rodriguez2023}. We use the interaction between the tip and the first atom of Ti as a control variable to transmit excitations from one side of the chain of Ti atoms to the other. Moreover, we employ this mechanism to initialize the spin chain in a suitable initial condition.  Our analysis considers the effects of short-range interactions, such as exchange, and longer-range interactions, such as dipolar interactions. We also consider the effect of tip anisotropy by including the magnetic anisotropy term in the Hamiltonian of the tip.

The paper is organised as follows. In Sec. \ref{smodel}, we introduce the model Hamiltonian of the spin chain coupled to the tip and outline the control mechanisms that will be explored throughout this work. Sec. \ref{sfree} presents the results for the simplest protocol where we  wait that autonomous, or non-controlled, evolution of the excitation achieves the desired outcome in a waiting time. Sec. \ref{sforced} is dedicated to the examination of an important point: improving dynamics through the use of the electric field between the tip and the surface. First, we explore the application of simple RF pulses and their tuning for achieving good quantum state transfer. Second, we propose the development of optimal pulse sequences to efficiently achieve desired outcomes in quite short times. Sec. \ref{sini} introduces a protocol for initializing the system before starting with the quantum state transfer process. Finally, in Sec. \ref{sc}, we summarise the most significant conclusions drawn from our work.

\begin{figure}[hbt]
\includegraphics[width=1.0\linewidth]{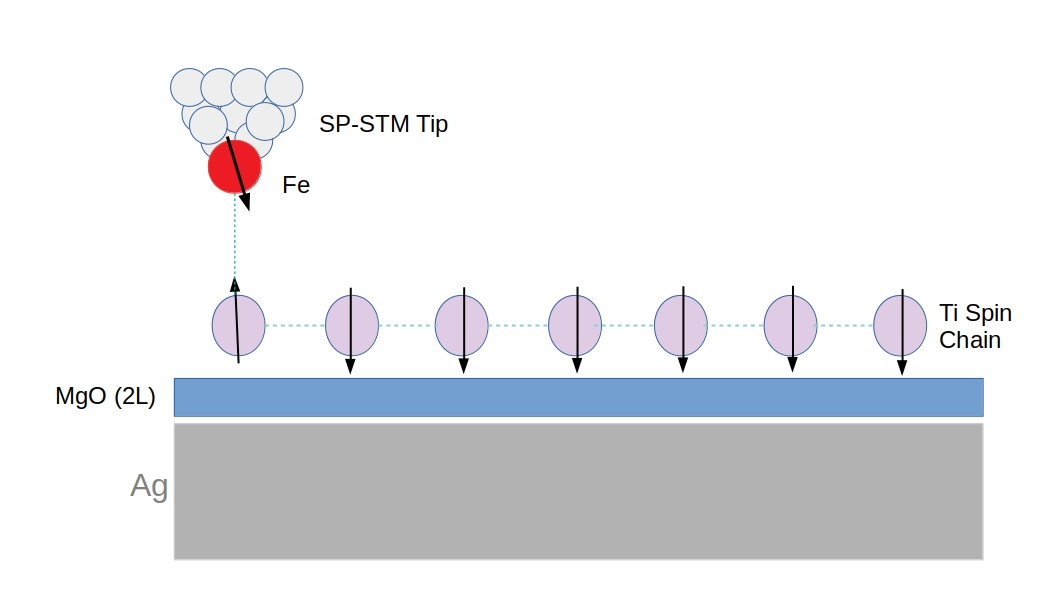}
\caption{\label{fig1} Sketch of a Ti spin chain in MgO/Ag and an SP-STM tip }
\end{figure}

\section{Model Hamiltonian and Controlled Quantum State Transmission}\label{smodel}

\subsection{Model Hamiltonian}

The study of the quantum dynamics in spin $1/2$ chains has garnered significant attention from numerous researchers over the past two decades. In particular, most of these works focused on the analysis of the efficient transmission of quantum states in Heisenberg and XX spin chains \cite{bose2007,coden2021,ferron2022,serra2022,serra2022b,zwick2011,zwick2011b,zwick2012}. This section deals with a more complicated and interesting problem, atomically engineered spin networks on surfaces. Notably, Titanium atoms deposited on Magnesium Oxide (MgO) have been observed to exhibit spin $1/2$ behavior\cite{yang2017,yang2019,willke2018b}. Furthermore, when these Ti atoms are arranged in more complex configurations, such as dimers\cite{yang2017}, their interactions are not only governed by exchange interactions but also influenced by dipolar interactions\cite{yang2017}. 

We assume that the Spin-Polarized STM (SP-STM) tip, typically composed of a piece of iridium coated with silver, housing approximately 1-5 iron (Fe) atoms at its apex \cite{willke2019}, can be effectively modeled by a single Fe atom. This Fe atom exhibits a spin $2$ behavior, and engages in a strong interaction with the first Ti atom of the chain. Moreover, when formulating the Hamiltonian for Fe, it is important to account for the anisotropy resulting from the underlying silver and iridium structure that provides its support \cite{mae1,mae2,mae3,mae4}.

The Hamiltonian of the SP-STM tip and the Ti atoms chain (see Fig. \ref{fig1}) system  written in terms of their effective spin degrees of freedom comprises two terms, $H=H_{1}+H_{2}$, where the first one models the interaction of spins with an external magnetic field and spin-spin interactions (exchange and dipole terms), and the second one models the local environment of the Fe atom at the tip. The first term is given by
\begin{eqnarray}\label{h1}\nonumber
H_{1}&=& \hbar^2\sum_{i<j} J(r_{ij})\, \mathbf{S}_i\cdot \mathbf{S}_{i+1} \\ \nonumber
        &+& \frac{\mu_{0}\mu_B^2}{4\pi}  \sum_{i<j} \frac{ g_i  g_j}{r_{ij}^3}[\mathbf{S}_i \cdot \mathbf{S}_{j}-3( \mathbf{S}_{i}\cdot \mathbf{\hat{n}}_{ij} ) (\mathbf{S}_{j}\cdot \mathbf{\hat{n}}_{ij} )]\\
        &-&\mu_B \sum_{i=0}^N g_i \mathbf{S}_i \cdot \mathbf{B}
        ,
\end{eqnarray}
\noindent where $\mu_0$ is the magnetic constant, $\mu_B$ is the Bohr magneton, the index value 0 refers to the spin of the Fe atom at the tip, $\mathbf{S}_0$, and the index values (1,...,N) refers to the Ti atoms of the spin chain on the MgO surface, $r_{ij}$ is the relative distance between the $i$-$j$ spins and $\hat{r}_{ij}$ its corresponding versor, $\mathbf{B}$ is the external magnetic field, $g_i$ is the effective gyromagnetic factor of the atomic specie at the site $i$, $g_0=2.7$ for Fe \cite{ferron2019,Rodriguez2023} and $g_i=1.8$ for Ti \cite{ferron2019}, and the $J_{ij}(r_{ij})$ are the exchange couplings defined as follows  
\begin{equation}
 J(r_{ij})=J_0\, e^{-(r_{ij}-r_0)/d_{ex}}
\end{equation}
\noindent where $J_0=0.0038$ meV, $r_0=8.64$\AA\, and $d_{ex}=0.4$\AA\, for the interaction between Ti spins and $J_0=64$ meV, $r_0=0$\AA\, and $d_{ex}=0.5$\AA\, for the interaction between Fe and Ti spins, are determined from previous experimental works \cite{yang2017,willke2019}. The Ti atoms were arranged in a linear arrangement whose interatomic distance was 8.64 \AA, which corresponds to locating the titanium over the oxygen atoms of the surface.

 The second term, that describes the SP-STM tip, is given by:
\begin{equation}\label{h2}
H_2=\hbar^2 \mathbf{S}_0\,\hat{{\cal D}}\,\mathbf{S}_0
\end{equation}
\noindent where $\hat{{\cal D}}=diag(0,0,D_{zz})$ is the second-rank local zero-field splitting tensor of the Fe atom because of the tip structure. The design of the STM tip is a relatively uncontrolled process, making it challenging to determine its characteristics \cite{Rodriguez2023}. We assume, for the rest of the work, a small value for the zero-field splitting tensor, $D_{zz}=0.05$ meV \cite{chen2023spinchain,mae1,mae4}. 

We perform exact diagonalization of the Hamiltonian. The  Hilbert space dimension is $2^N$ for the titanium chain and $5\cdot 2^N$ when the SP-STM tip is taken into account. We use the Quantum Toolbox in Python (QuTiP) \cite{QuTiP} to diagonalise the Hamiltonian and make posteriors quantum mechanical calculations. In the context of this study, we restrict our calculations to spin chains containing no more than 10 Ti atoms. We use the eigenstates and eigenvalues to  study the quantum dynamics in detail, as shown in the following subsections.

\subsection{Quantum State transfer: autonomous dynamical evolution}

The main idea of one of the simplest transfer protocol is to start with a spin 1/2 chain, here Ti atoms, where the first spin ($i=1$) is in an arbitrary state $|\psi_{0}\rangle$ and all other spins are in the down state $|{0}\rangle$, then the system evolves and after some time $T$ all the spins are in the state $|0\rangle$ except the last spin ($i=N$) which is in the state $|\psi_{0}\rangle$. To study the performance of the protocol note that: (1) the Hamiltonian in Eq. (\ref{h1}) does not commute with the $z$ component of the total spin of the system, which leads to the magnetization in the direction perpendicular to the surface not being conserved, and (2) because the protocol focuses on a subsystem of the total Hamiltonian, the chain of Ti, we need to study the dynamics after tracing over the Fe degree of freedom. These two characteristics distinguish our setting from the typical scenarios encountered in other studies without an ancillary spin. In particular, in addition to the state of the chain we need to specify the state of the Fe atom when we set the initial and target states and we need to use the full Hilbert space $5\cdot 2^N$ spanned by the basis states $|m_{0},\,m_1\,...\,m_N\rangle$ where $m_0\in(0,1,2,3,4)$ and $m_i\in(0,1)$ for $i\in(1,...,N)$. In ket notation for product states, the semicolon is used to split the parts corresponding to the Fe and Ti chain sectors of the Hilbert space.

In the case of the autonomous (uncontrolled) evolution, we set an initial state of the form $|\Psi_{0}\rangle=|\phi_{Fe},10...0\rangle$ where the $|\phi_{Fe}\rangle$ is the ground state of Fe atom in the absence of the chain of Ti atoms, and the state at time $t$ is given by $|\Psi(t)\rangle= \exp(-iHt/\hbar)|\Psi_{0}\rangle$. The figure of merit of the transfer protocol has to do with particular states of the Ti chain, irrespective of the Fe atom, so we trace over the Fe degree of freedom to construct the reduced density matrix that characterises the quantum state of the Ti chain. 
\begin{equation}
    \rho_{red}(t)=\textrm{Tr}_{Fe}\left(|\Psi(t)\rangle \langle \Psi(t)| \right).
\end{equation}

Finally, the figure of merit that we use to characterise the state transfer is the fidelity, refer also as Yield, defined by the function of the state of the system at time t through the reduced density matrix,

\begin{equation}\label{J1}
J_1[\Psi(t)]=\textrm{Tr}\left( \rho_{red}(t) |00...1\rangle \langle 00...1|\right).    
\end{equation}

\subsection{Quantum State Transfer: Driven evolution and  Optimal control of the state transfer}

Applying an electric field using the SP-STM tip induces a strain $\delta z$ of the bond between the Ti and the surface underneath \cite{lado2017}. In our case, Ti atoms can be located at the Oxygen site of the MgO bi-layer or at the bridge position between two oxygen atoms of the surface \cite{willke2018b,yang2017,yang2019}. The electric field leads to a modulation of the distance between the tip and the first atom of the chain \cite{lado2017,ferron2019,Rodriguez2023}. The
electric field exerted over a  Ti atom is given by, $E=V(t)/z_{tip}^{(0)}$, where $z_{tip}^{(0)}$ is the tip-Ti distance when there is no electric field and $V(t)$ is the electric potential difference between the tip and the surface. This field induces a force on an adatom, $F\simeq q V(t)/z_{tip}^{(0)}$ on account of its charge $q$. This force is compensated by a restoring elastic force $F=-k\delta z$. Then, we can evaluate how much a Ti atom of the spin chain is displaced from its equilibrium position \cite{lado2017} as:
\begin{equation}
        \delta z(t)\simeq\frac{q V(t)}{kz_{tip}^{(0)}}
        \label{piezo}
\end{equation}

\noindent where $q=1$ for the Ti atom and $k$ can be evaluated using DFT calculations \cite{lado2017,ferron2019} or using recent experimental data \cite{yang2017,yang2019}.  Eq. (\ref{piezo}) means that  the  Hamiltonian of our system, Eq. (\ref{h1}), depends on the electric field applied. 

We consider the electric field at the tunnel junction to be that produced by a parallel capacitor formed by the Ag tip and substrate as two metal electrodes \cite{wang2023qb,chu2020eftip}. In this scenario, each atom would feel the impact of the electric field. The specific influence on each atom varies depending on its position in the substrate. Here, we deal with Ti atoms located at  Oxygen sites, and then all atoms result equivalent. In this case, the chain undergoes a collective piezoelectric displacement, and there is no relative displacement between different Ti atoms. Thus, the main effect of the electric field is to change the relative distance between the tip and the first atom. A relative displacement between them  enables the possibility of remote control, as is discussed in \cite{soohion2023b} and \cite{phark2023double}. It can be interesting in the future to study spin chains with non equivalent atoms, such as TiB (bridge position)-TiO (atop Oxygen position) chains or Ti-Fe chains. 
 
We consider using the time-dependent \(z\)-component of the first Ti atom’s coordinate, influenced by a time-varying electric field exerted by the tip, as our actuator to drive the system dynamics. The time dependent behaviour of this actuator will be engineered, in conjunction with the magnetic field, to maximise the yield Eq. (\ref{J1}) of the transmission process. 

When the coordinate of one of the spins varies with time the functions that accompany the spin operators in the dipolar and exchange terms of the Hamiltonian also vary with time. If we consider the time variation of the coordinate of the spin $k$ and its particular influence on the spin $j$, the time dependent part of the Hamiltonian becomes $V=V_1+V_2+V_3+V_4+V_5+V_6$, where

\begin{eqnarray}
    V_1&=&  \left[J(r_{kj})+\frac{\mu_{0}(2\mu_{Ti})^2}{4\pi\hbar} \frac{(1-3 l^2)}{r_{kj}^3} \right]  S^x_{k}S^x_{j},\\ \nonumber
    V_2&=&  \left[J(r_{kj})+\frac{\mu_{0}(2\mu_{Ti})^2}{4\pi\hbar} \frac{(1-3m^2)}{r_{kj}^3} \right]  S^y_{k}S^y_{j},\\ \nonumber
    V_3&=&  \left[J(r_{kj})+\frac{\mu_{0}(2\mu_{Ti})^2}{4\pi\hbar} \frac{(1-3n^2)}{r_{kj}^3} \right]  S^z_{k}S^z_{j},\\ \nonumber
    V_4&=&  \left[-\frac{3\mu_{0}(2\mu_{Ti})^2}{4\pi\hbar} \frac{l\,m}{r_{kj}^3} \right]  (S^x_{k}S^y_{j}+S^y_{k}S^x_{j}),\\ \nonumber
    V_5&=&  \left[-\frac{3\mu_{0}(2\mu_{Ti})^2}{4\pi\hbar} \frac{l\,n}{r_{kj}^3} \right]  (S^x_{k}S^z_{j}+S^z_{k}S^x_{j}),\\ \nonumber
    V_6&=&  \left[-\frac{3\mu_{0}(2\mu_{Ti})^2}{4\pi\hbar} \frac{m\,n}{r_{kj}^3} \right]  (S^y_{k}S^z_{j}+S^z_{k}S^y_{j}),\\ \nonumber
\end{eqnarray}


\noindent and $l,m,n$ are the cosine directors of relative position between spin $k$ and $j$, and the explicit time dependence of the functions $r_{kj}(t),l(t),m(t),n(t)$ was omitted for clarity. It is worth mentioning that these six functions of time, for the pair $k,j$, are non linear on the actuator. As starting point we assume that the first Ti atom is moving with an harmonic time dependence due to a voltage $V(t)=V_{rf}\cos{\omega t}$. This time dependent problem was solved using routines available in QuTIP.

Besides studying the dynamics of the system using prescribed time-dependent electric fields, such as harmonic ones, we use Optimal Control Theory (OCT) [\onlinecite{Werschnik_2007}] to obtain the shape of time-dependent electric fields to maximise the yield functional. This variational problem, solved by the iterative algorithm of Krotov, includes the addition to the yield functional Eq. (\ref{J1}) of a functional that ensures the fulfillment of the Schr\"odinger equation, and another functional that ensures the smooth turn on and off of the electric field. 

This methodology aims to design the time-dependent amplitude $f(t)$ of a given operator $W$ so that a desired objective is achieved after time evolution. In our case, this objective is to maximize the yield after a predetermined total evolution time $T$. In its standard form, this theory assumes that the control function $f(t)$ appears linearly in the Hamiltonian through the term $f(t)W$. However, as mentioned earlier, due to the functional nature in which the spin coordinates appear in the dipolar and exchange interactions, our actuator enters the Hamiltonian in a nonlinear manner. Consequently, optimal control theory cannot, in principle, be used to design the electric fields required to drive the desired transition. Nevertheless, the experimental constraint on the possible displacements of the Ti atom allows us to linearize the Hamiltonian, enabling us to apply this theory. In this context, we calculate the electric fields with OCT ussing the linear approximation of the Hamiltonian and use these pulses to evolve the exact Hamiltonian, to report meaningful results in terms of the evolution of the exact Hamiltonian.

\section{Uncontrolled Transmission in a $Ti$ atom Spin Chain}\label{sfree}

 First, in this Section, we consider the scenario where the STM tip is either absent or positioned far enough away. Later on, we consider the autonomous evolution when the tip is close enough to the chain. 
 
 In the first case, we have a chain of Ti atoms of spin $1/2$ interacting through exchange and dipolar interactions and we prepare an excitation at the first atom and allow the system to evolve freely for a long enough time, and measure the fidelity given by Eq.(\ref{J1}) of the excitation reaching the final atom of the chain while the rest of the spins in the chain remaining in the $|0\rangle$ state. In contrast with the case studied by Bose in [\onlinecite{bose2007}] no symmetry prevent the free dynamics to abandon the total z-magnetization subspace where the initial and final state lives because of the presence of the dipolar interactions terms in the Hamiltonian in Eq. (\ref{h1}). 

\begin{figure}[hbt]
\includegraphics[width=0.85\linewidth]{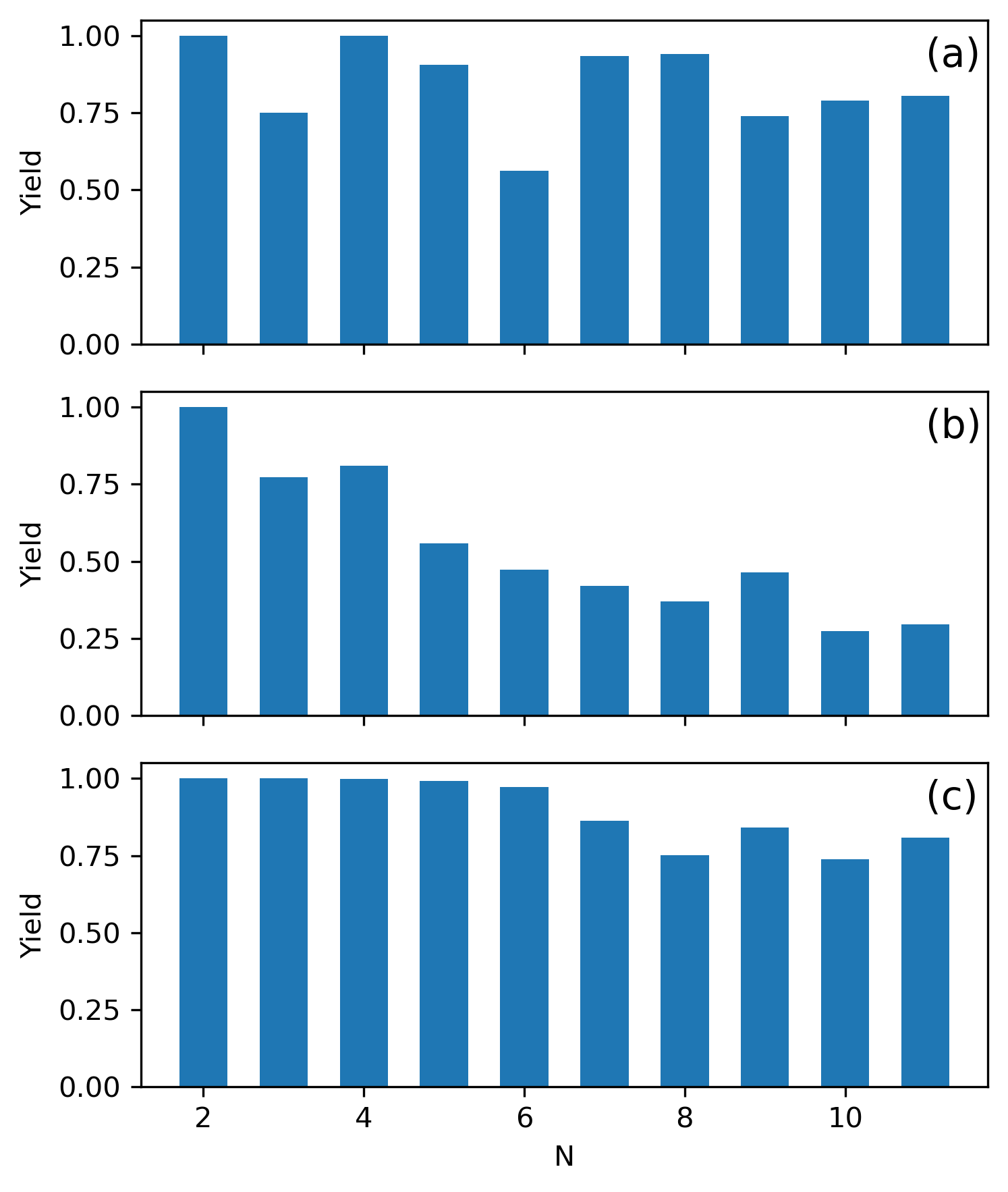}
\caption{\label{figbs} Yield for different Ti chains without the interaction with the SP-STM tip and considering: (a) only the Heisenberg term in Eq. (\ref{h1}) (first term), (b) the Dipolar and Heisenberg terms in Eq. (\ref{h1}) with no magnetic field (first and second term). (c)  the Dipolar and Heisenberg terms in Eq. (\ref{h1}) with a magnetic field.}
\end{figure}

In Fig. \ref{figbs}, we plot the maximum yield obtained after waiting up to $T_{max}=200$ ns, a time close to the decoherence time for this kind of system [\onlinecite{Baumann2015}].
In Fig. \ref{figbs} (a) the maximum yield is shown for the exchange-only interaction between spins of the chain of Ti atoms obtained by turning off the dipolar terms in the Hamiltonian. This maximum yield is unaffected by the presence of an overall magnetic field and, as expected, the outcome is the same as the one obtained by Bose \cite{bose2007}. When we introduce the dipolar interaction, however, the picture is quite different, as depicted in Fig. \ref{figbs} (b). In panel (b),  a noteworthy and abrupt diminution of the yield is observed. An interesting situation occurs when we turn on the magnetic field, as shown in panel (c), and the yield  improves in reference to that of the Heisenberg spin chain with just exchange interactions. 

\begin{figure}[hbt]
\includegraphics[width=0.80\linewidth]{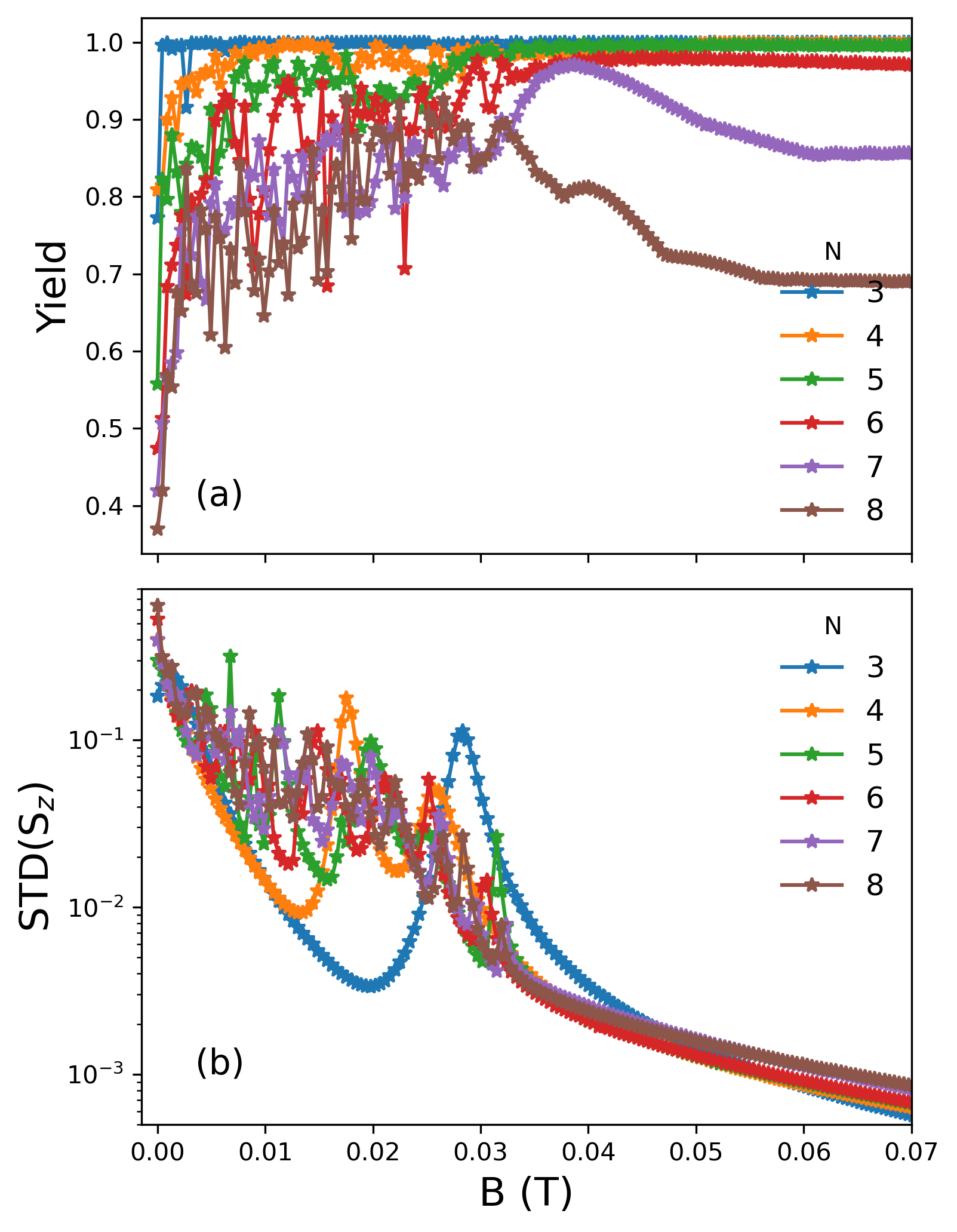}
\caption{\label{figB}  Effect of an external magnetic field perpendicular to the surface in the quantum state transmission in Ti-spin chains. (a) Yield of different Ti spin chains as a function of the external magnetic field. (b) Standard deviation of the total $S_z$ as a function of the external magnetic field for different chains. }
\end{figure}

The dipolar interaction does not preserve the initial magnetization which is  problematic because we are interested in a transmission between two states that belong to the same total magnetization subspace, which is now not automatically guaranteed as would have been the case if the interaction was that of Heisenberg. This incursion of the state outside the relevant subspace during it evolution is interpreted as a leakage to sub-spaces with more than one excitation and results in considerably less efficient dynamics in terms of the yield. Figure  \ref{figbs}(b) clearly illustrates this. However, we can control this leakage turning on an external magnetic field as illustrated in Fig. \ref{figB}(b). In that figure, for different values of the external magnetic field, we plot the standard deviation of the expected value of the $z$ component of the total spin, $\langle\Psi(t)|S_z|\Psi(t)\rangle$, during the dynamics of the chain starting in the state $|10...0\rangle$ and observed on an interval of 200ns. It can be appreciated that still for low magnetic fields the state mostly remains in the relevant subspace and its expected value, as a function of time, exhibits an almost constant behaviour. Beyond $30$ mT, there is no discernible leakage into spaces with more than one excitation. This explains the phenomenon observed in Fig. \ref{figbs}(c).

It is well known that, depending on the relative strength of the dipolar interaction with respect to the interaction with the magnetic field, the dipolar terms parallel to the surface can be almost negligible. Nevertheless, we calculate the dynamical evolution of the quantum states using all the terms in Eq.~(\ref{h1}) in the whole Hilbert space, irrespective of the magnetic field strength. 

In Fig. \ref{figB} (a), we plot the yield in chains with different number of Ti atoms, as a function of the magnetic field, paying attention to small magnetic field values. It is evident that for small magnetic fields, the outcomes exhibit a rapidly fluctuating behaviour of the yield with respect to the magnetic field strength. This behaviour is in agreement with what we observe in Fig. \ref{figB} (b); as the dynamical evolution includes visits to different total magnetization sub-spaces, the magnetization undergoes fluctuations, making difficult the stabilization of the yield. However, as we increase the magnetic field strength, the variability of the expected value of $S_z$ decreases. Two regimes appear, for short enough chains the yield reaches a maximum value and this value does not change despite increasing values of the magnetic field strength. For larger chains, after reaching a maximum value the increasing of the magnetic field strength diminishes the value of the yield. 

\begin{figure}[hbt]
\includegraphics[width=0.85\linewidth]{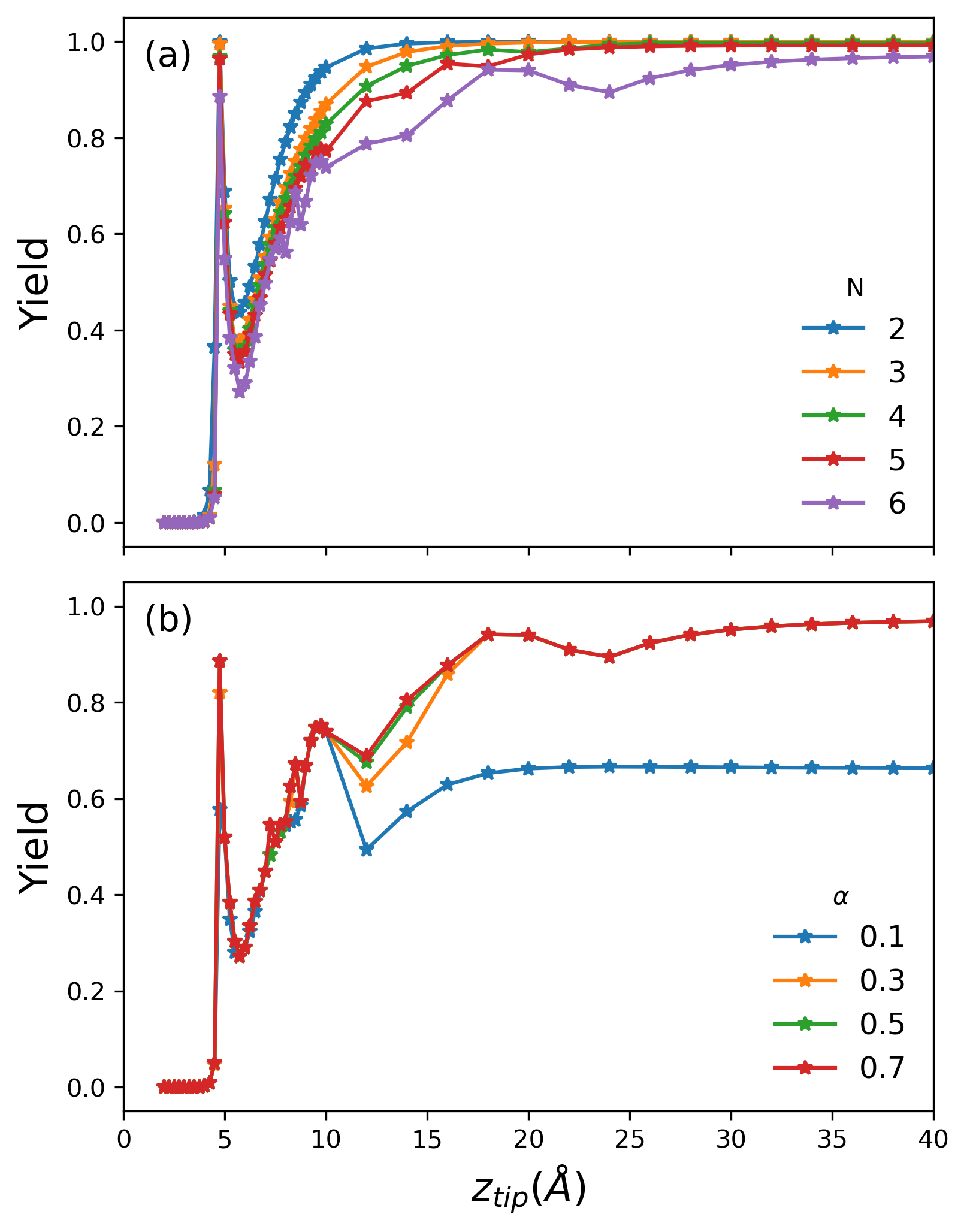}
\caption{\label{fignotin} Yield as a function of the tip atom distance. (a) Yield for different length chains. (b) Yield for $N=6$ chain and different waiting times $T=\alpha T_{max}$.}
\end{figure}

If we aim to perform some measurements during the free evolution or seek to improve the efficiency of the transmission, it is necessary to bring the SP-STM tip closer. When the tip is close enough to the first Ti atom, it influences the excitation dynamics through the chain. In Fig. \ref{fignotin}, we once again show calculations of the maximum yield achieved for time intervals shorter than $T_{max}=200$ ns, but now with the presence of the tip, so the Fe atom term now enters into the Hamiltonian. We  prepare the system in the state $|\phi_{Fe},10...0\rangle$ and study the dynamics of the reduced density matrix of the spin chain to asses the probability of obtaining an excitation in the last atom of the chain, with the rest of the Ti atoms in its ground state and ignoring all degrees of freedom of Fe (see Eq. (\ref{J1})). 

Fig. \ref{fignotin} (a) depicts the maximum yield for chains with different number of atoms, as a function of the tip-atom distance $z_{tip}$. We observe that placing the tip too close destroys the free propagation of the excitation through the channel. For distances lower than $4$ \AA\,  the excitation does not reach the last site. When the tip lies at a long enough distance from the first atom, the transmission yield resembles the one observed in an isolated Ti chain. There are two noteworthy features that emerge from a detailed observation of Fig. \ref{fignotin}. 

Firstly, the distance at which the influence of the tip is negligible depends on the size of the chain. The distance between successive eigenvalues diminishes for increasing chain lengths so, a perturbation of a given strength might modify the dynamics of the chain excitations, unlike other characteristics, more strongly for larger chains. For the effect of the tip to be negligible, it is necessary to move away the tip much further than it could be expected from experimental data in standard ESR-STM experiments performed in small atoms arrangements.  

Secondly, an interesting feature observed in Fig. \ref{fignotin} is the peak observed at $4.75$ \AA\,. At this tip-atom distance, we appreciate a near-complete restoration of the excitation transmission of the isolated chain. This phenomenon arises because, at the No Tip Influence point (NOTIN point) \cite{Rodriguez2023,seifert2021}, the competition between the dipole interaction and exchange interaction almost cancels the tip influence on the first Ti atom. This region seems to be an ideal working region, allowing proximity to the system without compromising excitation transmission. However, this peak is extremely sharp, making experimental operation complicated. An alternative option is to operate with the tip placed at distances of $10$ \AA\, or $15$ \AA\,, where the chain still interacts with the tip but transmission remains reasonably high.

Other critical aspect to consider is the time required to achieve good yields. For the protocol to be considered efficient and practical, it is necessary that transmission times remain short compared with the decoherence time. By reducing the maximum waiting time in our calculations, we observe that when dealing with chains with $N\leq 5$, there are no important differences in the outcomes up to $T_{max}=20$ ns. Nevertheless, with larger chains, a more detailed analysis is needed. In Fig. \ref{fignotin} (b), we observe that for some positions of the tip, we need waiting times exceeding $60$ ns for a $N=6$ spin chain. It is worth noting that for very short times, around $20$ ns, even the isolated Ti chain exhibits poor transmission efficiency.

The waiting time necessary to obtain an excellent quantum state transfer in homogeneous and nonhomogeneous chains with an autonomous time evolution depends on a nonlinear fashion with the chain length. It also depends on the required value asked for the yield, the closer the yield value to the unity, the larger the waiting time. These dependencies explain the behaviour observed in Fig.~\ref{fignotin} (a) and (b).

\section{Driving the transmission using an STM}\label{sforced}

Recent experiments have shown the possibility of manipulating the quantum state of an atom using the electric field generated between the STM tip and a surface
\cite{yang2018,soohion2023b,sooyhon2023,willke2019,willke2019b,willke2018}. Our goal in this Section is to use the electric field of the tip to control quantum state transmission through Ti atom chains. We propose two methods to achieve this. The first method involves exciting the system with radio frequencies (RF) and observing its evolution. The second and more technically challenging method involves optimizing the strength and time dependency of the applied electric field to achieve efficient state transmission within a chosen waiting time.

\subsection{RF driving}

Our main goal is to enhance the quantum state transmission performance using a time-dependent electric field, similar to most ESR-STM experiments. Unlike conventional methods which consider a range of frequencies to detect resonances, we aim to find the optimal frequency to maximize the transmission efficiency without paying attention to the resonance frequencies. 

We introduce a time-dependent potential 

\begin{equation}
V(t)=V_{rf}\cos{(\omega t)}    
\end{equation}

\noindent that produces a piezoelectric displacement $\delta z(t)$ of the Ti atom as shows Eq. (\ref{piezo}). In the case of Ti at MgO, we have $q=1 e$ and a coupling constant $k$ that we need to estimate. Recent experiments\cite{yang2019} show that the displacement of the Ti atom at the oxygen site, when applying a constant voltage of $10$ mV with the tip $4.3$ \AA $\,$  above the atom is close to $\delta z=3$ pm. Using Eq. (\ref{piezo}), we obtain $k=0.08$ $eV/$ \AA$^2$. 

We solve the time-dependent Schr\"odinger equation for a Hamiltonian with explicit time-dependent terms. The temporal dependence arises from the displacement of the first Ti atom because of the influence of the external applied electric field. Specifically, the explicit time-dependence is present in the interaction terms between this first site and the tip.

\begin{figure}[hbt]
\includegraphics[width=0.9\linewidth]{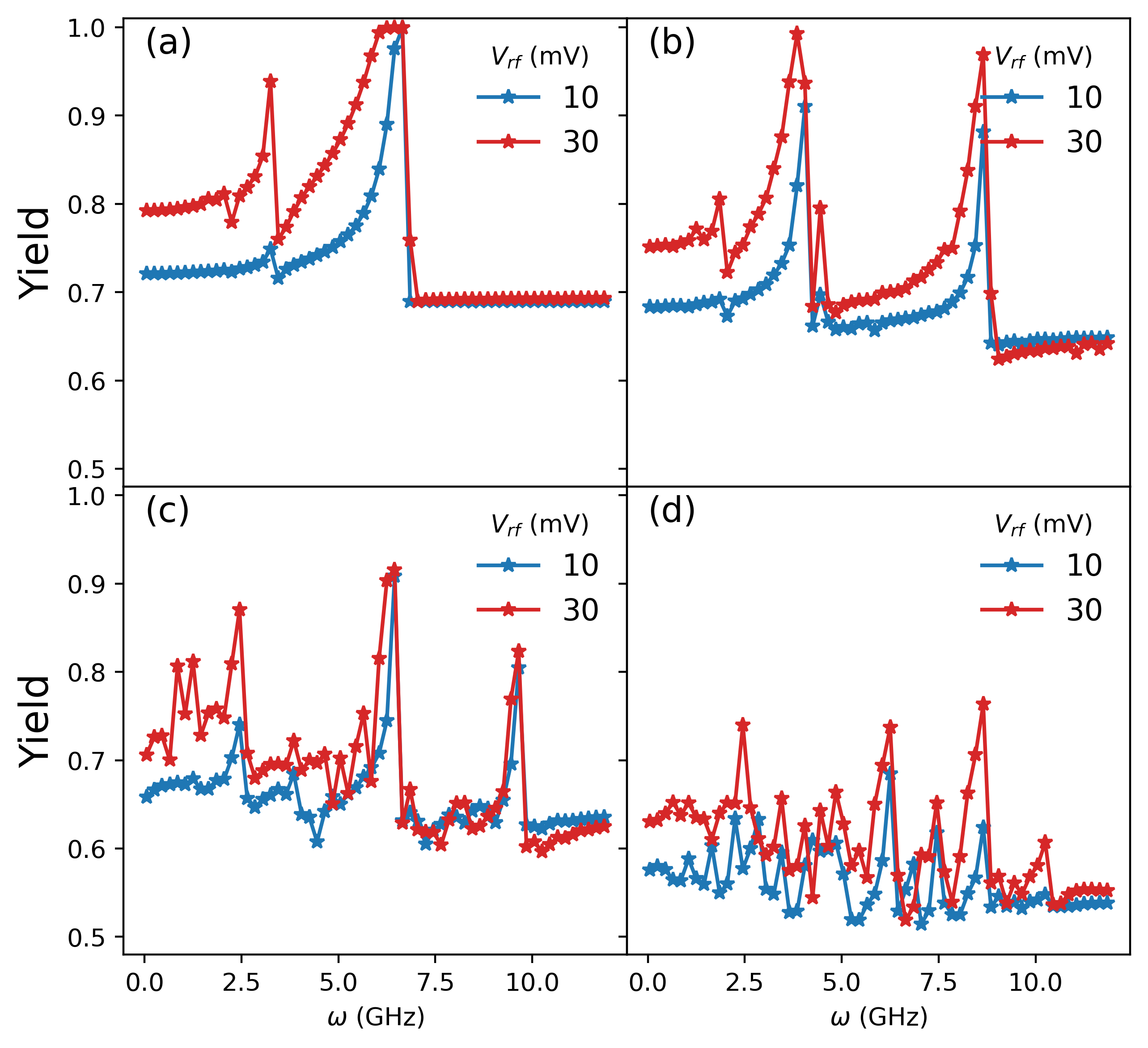}
\caption{Yield as a function of the driving frequency for different amplitude voltages. (a)N=2, (b)N=3, (c)N=4, and (d) N=6. \label{driveRF-w}  }
\end{figure}

In Fig. \ref{driveRF-w}, we see how it is possible to significantly improve the transmission by tuning the electric field frequency. The protocol consists of sweeping the driving frequency at a fixed amplitude, a procedure similar to that used in most STM-ESR experiments, and fixing the waiting time at $200$ ns. For each frequency, as in the free evolution, we choose the most favorable outcome in the interval defined by the waiting time. It is impressive that, for small chains, the improvement in efficiency is enormous. As depicted in Figs. \ref{driveRF-w}(a) and \ref{driveRF-w}(b), at specific frequency values, transmission rates improve from $60\%$ or $70\%$ to nearly $100\%$ in the cases $N=2$ and $N=3$. The situation becomes more intricate as we extend the length of the chains. As we observe in Fig. \ref{driveRF-w}(c), for $N=4$, the periodic driving enhances the transmission from $55\%$ to more than $90\%$ while in Fig. \ref{driveRF-w}(d), for $N=6$,  the driving takes the yield from $55\%$ to almost $75\%$. For longer chains, achieving fairly good yields values becomes more difficult, requiring extreme fine-tuning of the frequencies (see Fig. \ref{driveRF-w}(d)).

\begin{figure}[hbt]
\includegraphics[width=0.9\linewidth]{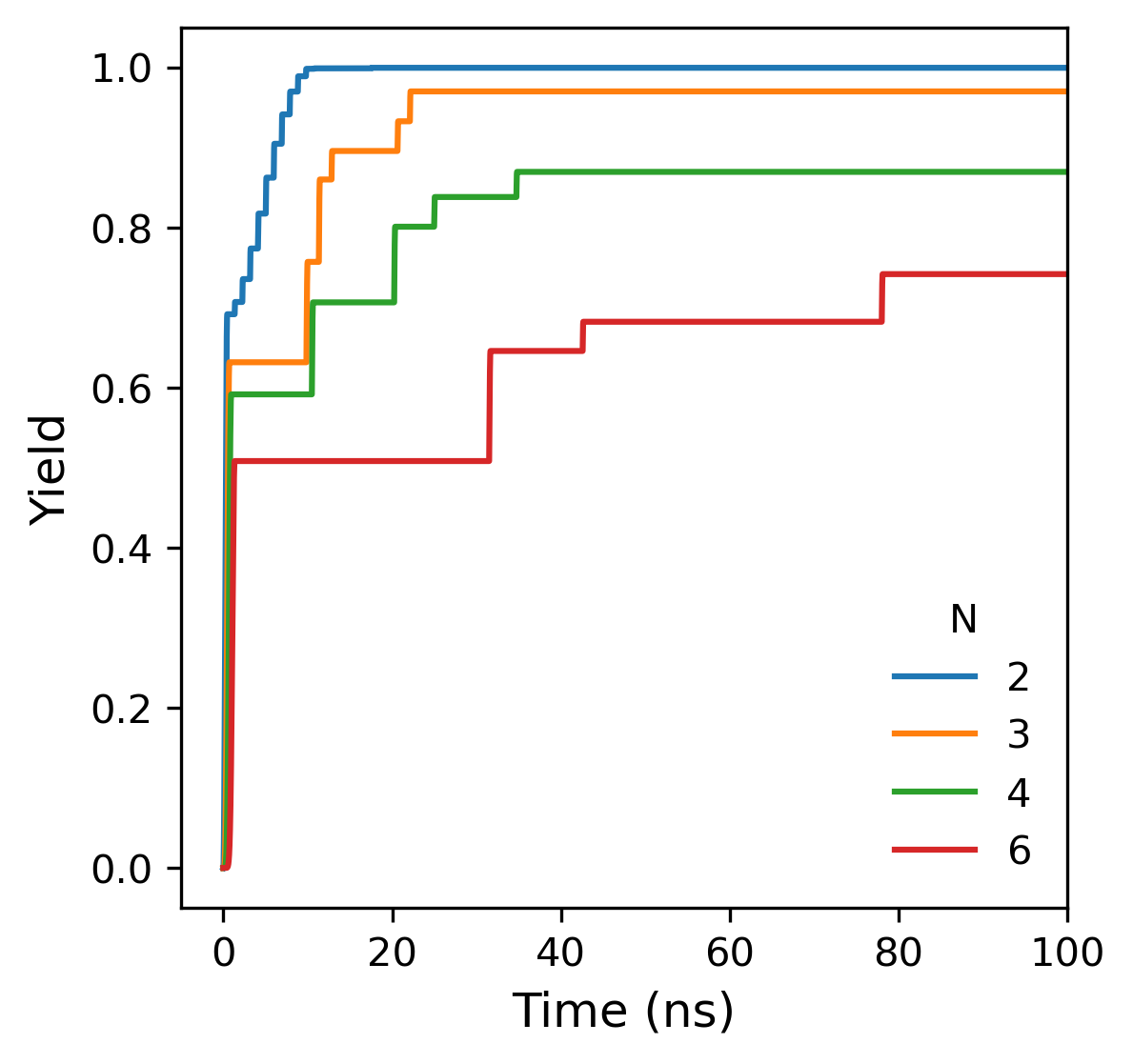}
\caption{\label{driveRF-time} Yield for the optimal frequency obtained in Fig. \ref{driveRF-w} as a function of the waiting time for different Ti chains.}
\end{figure}

The waiting time plays an important role. Very long waiting times spoil the transmission  due to decoherence. In the same way, very short waiting times prevents  the driving from improving the efficiency of the transmission. In Fig.~\ref{driveRF-time}, we observe the yield for the optimal frequency as a function of the waiting time. As expected, longer waiting times are needed as we deal with longer chains. It is notable that, after a given time, increasing the waiting time is unnecessary. The time at which the yield stabilises also depends on the size of the chains. It becomes evident that for waiting times exceeding 40 ns, we have achieved the desired regime in all the analysed chains. In the case of N=6, there might be the temptation to extend the waiting time by another 40 ns to improve the transmission marginally, but this strategy is not advisable in actual implementations owed to the presence of decoherence. It is important to note that the frequency range explored in this work (see Fig. \ref{driveRF-w}) is experimentally accessible with state-of-the-art equipment employed in ESR-STM experiments.

\subsection{OCT designed driving}

\begin{figure}[hbt]
\includegraphics[width=0.9\linewidth]{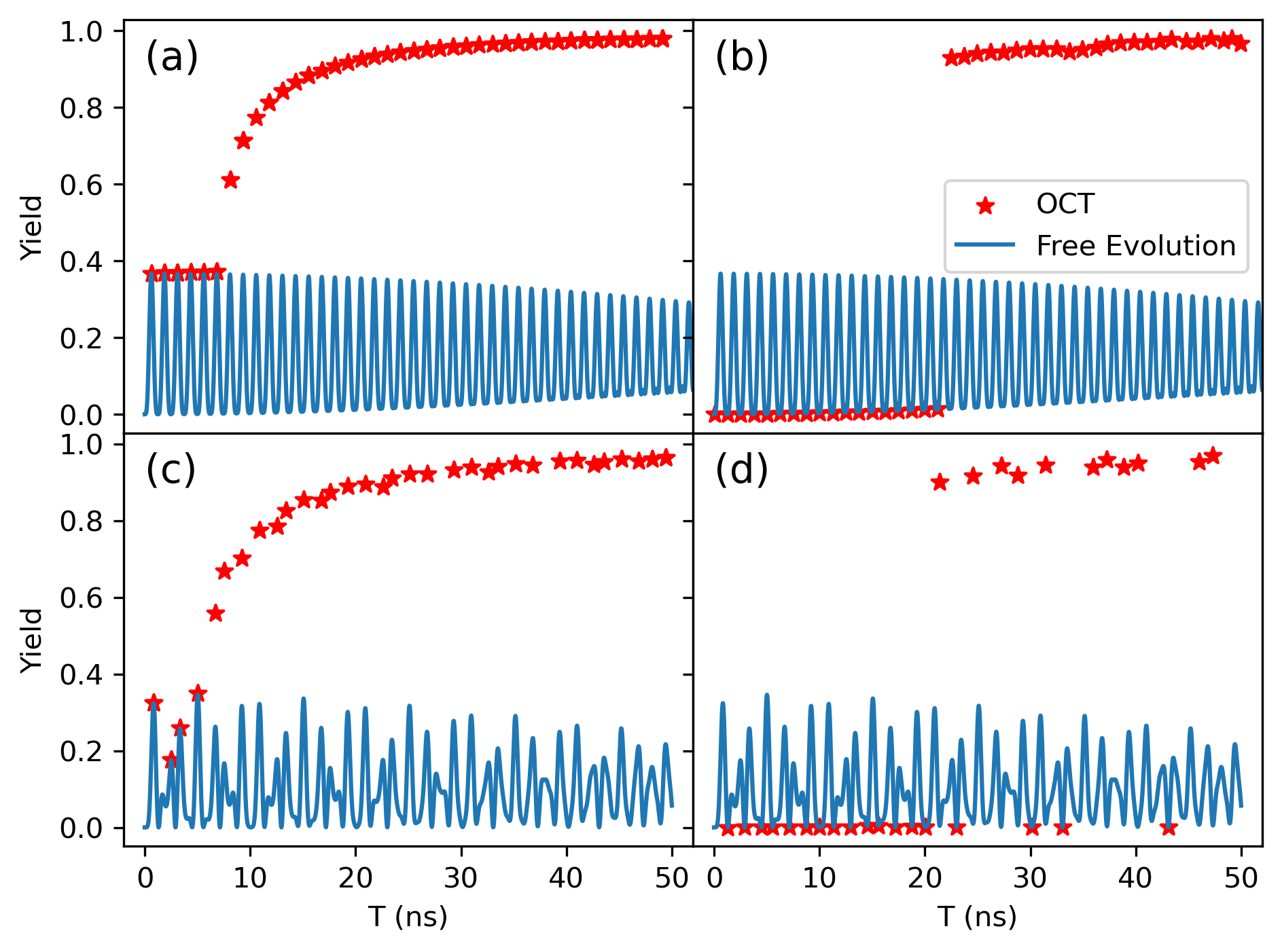}
\caption{\label{foct1} Free evolution (blue lines) and yield obtained using OCT pulses at fixed times (red stars) for a Ti spin chain of length $N$ with the STM tip located at $z_{tip}=5.75$ \AA. The left panels show times chosen in the maximum yields attained in the autonomous evolution and the right panels show the same for the minimum yields attained for the autonomous evolution. The chain lengths considered are $N=3$ and $N=4$ for the upper and lower panels, respectively.}
\end{figure}

\begin{figure}[hbt]
\includegraphics[width=0.9\linewidth]{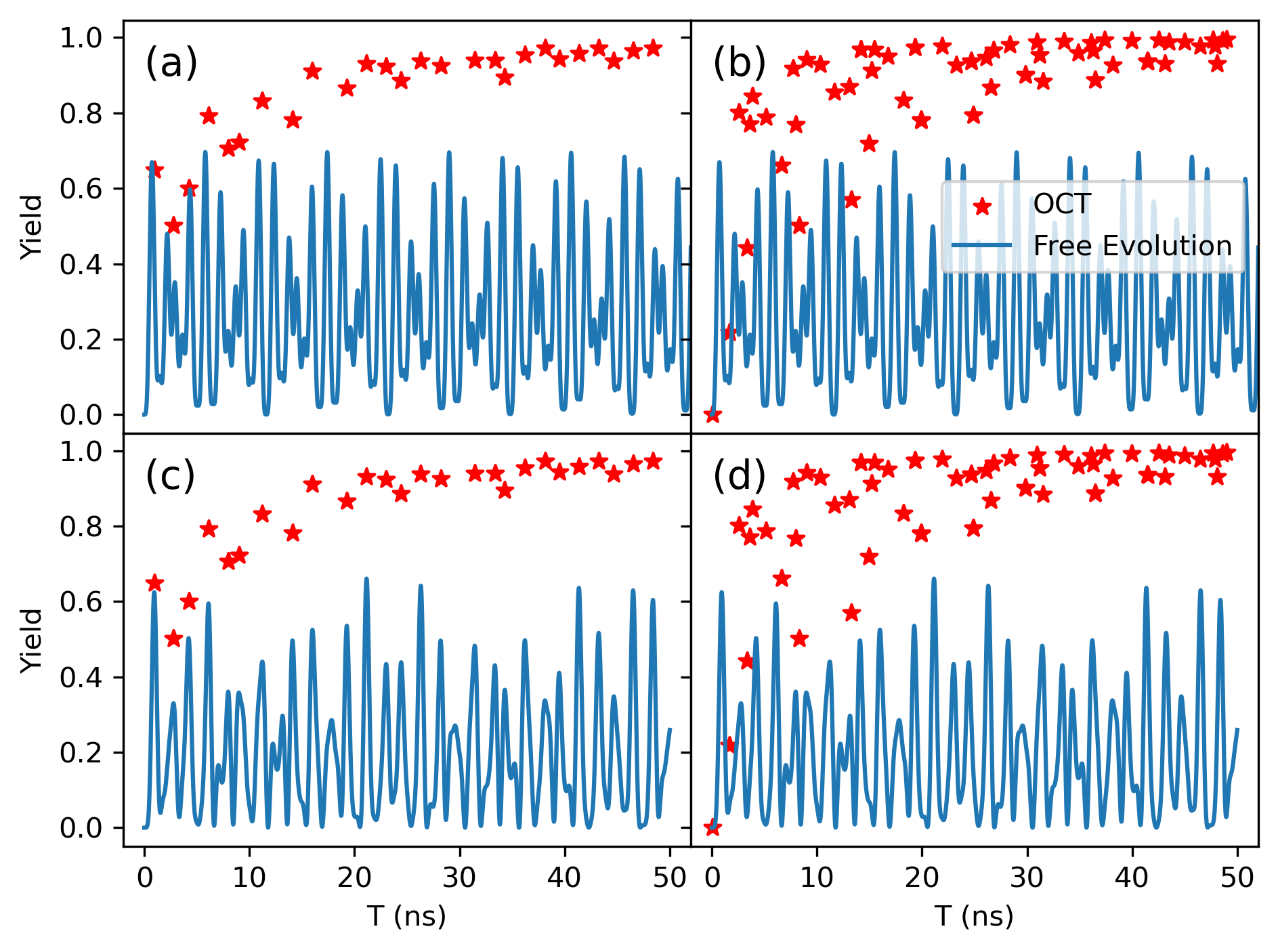}
\caption{\label{foct2}Free evolution (blue lines) and yield obtained using OCT pulses at fixed times (red stars) for a Ti spin chain of length $N$ with the STM tip located at $z_{tip}=8.0$ \AA. The right panels show times chosen in the maximum yields reached in the free evolution and the right panels show the same for the minimum yields reached for the free evolution. The chain lengths considered are $N=3$ and $N=4$ for the upper and lower panels, respectively. }
\end{figure}

In this section, we use Optimal Control Theory to design electric control fields which enhance the transmission efficiency, by modulating the interaction between the Fe atom and the first Ti atom through adjustments in the $z$ component of the position of the first Ti atom. Figures \ref{foct1} and \ref{foct2} show the yield obtained applying the OCT, for two distinct tip positions alongside two Ti chains. The blue lines depict results obtained through free evolution, while the red stars represent the OCT outcomes for a predetermined time. The selection of these operation times is such that it coincides either with the maxima (in the left panels) or with the minima  (in the right panels) obtained during autonomous evolution.

In Figs. \ref{foct1} and \ref{foct2} , it is observed that a minimum time is required for Optimal Control Theory (OCT) to enhance the results. For times shorter than this critical time, the autonomous evolution seems to be the best option. In Fig. \ref{foct1}, we show results for the tip close to the atom, where free evolution shows poor performance. Notably, when we select times corresponding to the maxima of the free evolution, improvements in transmission with OCT begin for times lower than $10$ ns. Immediately after the critical time, transmission efficiency jumps from $30$-$40\%$ to $60\%$. For times exceeding $30$ ns, transmission efficiency can easily surpass $90\%$ (see left panels of Fig. \ref{foct1}). However, selecting times corresponding to the minima in free evolution complicates the objectives. Despite this, transmission is still significantly improved, although this choice results in a much longer critical time, besides OCT does not always converge to a reasonable solution. Nevertheless, for certain times the maxima correspond to quantum state transfer of very high quality. The highest yields values for each panel are (a) 0.979, (b) 0.979, (c) 0.963, and (d) 0.969.

Figure \ref{foct2} presents a similar analysis, but the tip is located further away than in Fig.~\ref{foct1}. At this tip-atom distance, the free evolution is more efficient. The critical time for OCT to improve transmission is shorter than in the previous case, but the behaviour of the yields obtained using OCT as a function of time is more convoluted. While the transmission yield can still reach values larger than $90\%$, it varies significantly with the operation time, especially when using  times corresponding to the minima of the autonomous evolution. Interestingly, in this case, times corresponding to the minima of the temporal evolution can achieve very efficient transmissions for certain specific operation times. However, the right panel of Fig. \ref{foct2} shows considerable dispersion in the achieved yields, indicating that  designing an efficient protocol becomes challenging. In the left panel, for operation times corresponding to the maxima in the free evolution, transmissions exceeding $90\%$ are observed with slightly less dispersion. The OCT results show that despite the variations observed in the maxima heights, for certain times the maxima correspond to quantum state transfer of very high quality. The highest yields values for each panel are (a) 0.973, (b) 0.996, (c) 0.973, and (d) 0.996. 

The results shown in both Figures, \ref{foct1} and \ref{foct2}, rise some interesting points. While it is clear that if what is required is the highest possible yields, then the tip of the STM must be placed at longer distances from the first Ti atom. But this ask for a very fine tuning of the instant in which the information could be retrieved, otherwise the fluctuations observed in the height of the yields peaks could spoil the OCT procedure.  

Next, we analyse the time behaviour of the control pulses designed using OCT. Numerical algorithms implementing OCT always produce a control pulse, but its utility in an actual implementation is not guaranteed. Introducing physical constraints to limit the frequencies and strength of the pulses is not simple, therefore many times a direct observation of the pulses is mandatory.

\begin{figure}[hbt]
\includegraphics[width=0.9\linewidth]{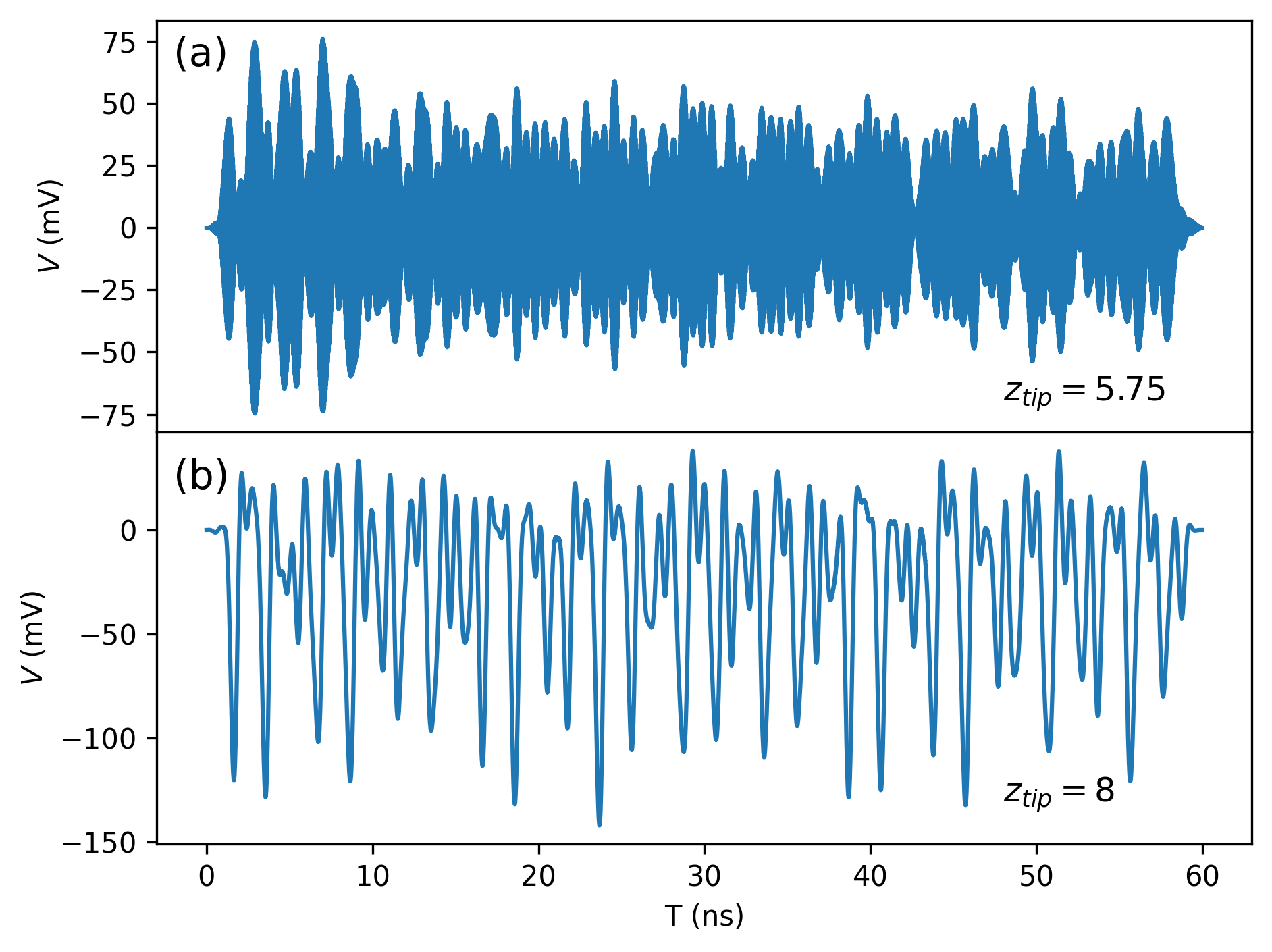}
\caption{\label{pulse} OCT pulses for a Ti spin chain of length $N=4$ with the STM tip located at $z_{tip}=8.0$ \AA $\,$ (lower panel) and at $z_{tip}=5.75$ \AA $\,$ (upper panel). The duration of both pulses is $T=60$ ns and the yield reached are $0.987$ and $0.996$ respectively.  }
\end{figure}

Figure \ref{pulse} displays the pulses for $T=60$ ns for a four Ti atom chain and two different tip positions. The achieved yields are 0.987 for $z=5.75$ \AA, and 0.996 for $z=8$\AA. Note that for these tip positions the autonomous evolution renders very different values of the yield, see Fig.~\ref{fignotin}, while the driven time evolution renders values that differ in $1\%$. While OCT significantly improves the yield obtained with respect to  the one obtained from the autonomous transmission in both cases, the pulse obtained when the tip is further away from the surface is much simpler. The Fourier spectrum (not shown) indicates that the frequency spectrum in panel (a) is more complex and the main frequencies an order of magnitude higher than those in panel (b). It is worth to mention that both scenarios are challenging for the OCT algorithm since we enforce a specific high-yield time. Nevertheless, both control scenarios feature smooth turn on and off and experimentally implementable frequencies.

We present results concerning only linear arrays with successive atoms separated by the same constant distance. These restrictions do not affect, in principle, the ability of the OCT to find control pulses that improve the transmission.


\section{Preparation of the initial state and a complete protocol for quantum state transmission}\label{sini}

In this section we apply the OCT to prepare a particular state on the first Ti atom. The transmission protocol of arbitrary states of one qubit implies that at least a set of distinguishable quantum states can be prepared as required by the information to be transmitted. Ideally, this set comprises the whole pure state space. In an actual implementation there are severe restrictions. We explore the preparation of quantum states within the restrictions imposed by the use of a STM. 

Our preparation protocol involves two steps. The first step implies cooling the system and applying a magnetic field to set the chain in its ground state (all the spins down). The second step requires flipping the first spin of the chain. For this procedure, the magnetic field is turned off and the electric field of the tip drives the chain until the desired initial state is obtained.

We found that to prepare the chain state $|1000\ldots\rangle$, the best results correspond to a tip-atom distance of $5$ \AA, or more generally to the right of the NOTIN point but close to it. In this case the preparation and subsequent control of the transmission are guaranteed. Moving away the tip from the NOTIN point guarantees a more controlled transmission but the preparation is achieved with a very low fidelity. Setting the tip-atom distance to values smaller than the one of the NOTIN point results in an excellent preparation of the initial state, but as seen above this precludes the transmission of any information. 

Using OCT we prepare the initial state $|1000\ldots\rangle$ with a yield up to $0.97$, for a tip distance of $5$\AA, and a preparation time of $40$ ns, which is a short time that, when  added to the transmission time, does not result in an time too long to implement the whole transmission protocol when compared to the possible decoherence times. 

Other initial states could be prepared in shorter times, or of the order, than the state $|1000\ldots\rangle$. The procedure to deal with imperfect preparation of the initial state, and faulty transmission has been discussed elsewhere, see Reference~\cite{serra2022b}. 

\section{Summary and Conclusions}
\label{sc}

In this work, we have explored the feasibility of controlled quantum state transfer in a chain of spin-1/2 titanium atoms adsorbed in a MgO surface using the electric field produced by a Scanning Tunneling Microscope in addition with a constant magnetic field perpendicular to the surface. The employment of Optimal Control Theory significantly enhances the efficiency of quantum state transmission in the Ti atom spin chain with respect to the waiting for the best yield on both, the autonomous evolution and the harmonic forced dynamics. The complexity of these pulses varied with the tip's distance from the chain, with more intricate frequency spectra observed for shorter distances. These pulses are experimentally feasible, suggesting that our OCT-based protocol can be readily implemented in practical STM setups. The study highlighted the challenges in designing efficient transmission protocols due to precise timing required for the OCT algorithm on short times. 

We successfully prepared the initial state $\vert 1000 \ldots \rangle$ using OCT, achieving a yield of up to 0.97 for a tip distance of 5 \AA \ and a preparation time of 40 ns. This preparation method is efficient and integrates well with the overall transmission protocol, minimizing additional implementation time. The results of this study demonstrate the potential of STM-controlled spin chains for quantum information processing. The high-fidelity state transfer and the feasibility of control pulse implementation underscore the practical relevance of this approach. As a final conclusion, we think that the present work contains enough evidence that quantum state transfer is implementable in atomic chains of adatoms with state-of-the-art ESR-STM experimental devices. In summary, our findings confirm the viability of using STM-induced electric fields for high-fidelity quantum state transfer in spin-1/2 chains. The integration of OCT provides a robust framework for optimizing these processes, paving the way for advancements in atomic-scale quantum information technologies.

\section*{Acknowledgements}

OO acknowledges partial financial support from CONICET (PIP 11220210100787CO) and Secyt-UNC (33620230100363CB), and the warm hospitality at FACENA-UNNE. AF and DAC acknowledge funding from 
CONICET (PUE22920170100089CO and 
PIP11220200100170), AF acknowledges partial financial support from 
ANPCyT (PICT2019-0654)

\bibliography{biblio}{}

\end{document}